# Exact Solution for a Black Hole Embedded in a Nonstatic Dust-filled Universe

E. Kopteva[1], P. Jalůvková[1,2], I. Bormotova[1,2], and Z. Stuchlík[1]
[1] Institute of Physics, and Research Centre of Theoretical Physics and Astrophysics, Faculty of Philosophy and Science in Opava, Silesian University in Opava, 746 01 Opava, Czech Republic; olena.kopteva@gmail.com
[2] Bogoliubov Laboratory of Theoretical Physics, Joint Institute for Nuclear Research 141980 Dubna, Russia

## Abstract

An exact solution of the Lemaître–Tolman–Bondi class is investigated as a possible model of the Schwarzschild-like black hole embedded in a nonstatic dust-filled universe for the three types of spatial curvature. The solution is obtained in comoving coordinates by means of the mass function method. It is shown that the central part of space contains a Schwarzschild-like black hole. The R–T structure of the resulting spacetime is built. It is shown that the solution includes both the Schwarzschild and Friedmann solutions as its natural limits. The geodesic equations for test particles are analyzed. The particle observable velocities are found. The trajectories of the test particles are built from the point of view of both comoving and distant observers. For the distant observer, the results coincide with the Schwarzschild picture within a second-order accuracy near the symmetry center.

## 1. Introduction

An interest in cosmological black holes still prevails due to their realism and importance in numerous fundamental problems in cosmology and relativistic astrophysics (Faraoni & Jacques 2007; Tada & Yokoyama 2015). The term "cosmological" black holes usually refers to black holes immersed in a certain cosmological medium that fills the whole universe.

In this regard, the most frequently mentioned in the literature is the McVittie solution (McVittie 1933), generally used for the description of the cosmological black holes (Nolan 1998; Faraoni & Jacques 2007; Lake & Abdelgader 2011; da Silva et al. 2013; Nolan 2014), although it was shown that this solution and its modifications cannot describe the black hole in a homogeneous dust background (Korkina & Kopteva 2012).

On the other hand, the Einstein–Straus (ES) model (Einstein & Straus 1945) was commonly used to treat a universe with an embedded black hole for many years (see, e.g., Thakurta 1981; Stuchlik 1983, 1984; Bonnor 2000; Mars 2001; Sultana & Dyer 2005 and references therein). However, it was shown that ES configurations are unstable with respect to radial perturbations (Plebanski & Krasinski 2006), and the empty vacuoles in them should have rather big diameters when applied to real objects in the universe (Bonnor 1996). Besides, it was shown in Korkina & Kopteva (2012) that some generalizations of the ES solution, like those in Thakurta (1981) and Sultana & Dyer (2005), turn out to be modified McVittie solutions.

An alternative (and free of the mentioned lacks) approach to describe the black holes within their cosmological surroundings is to use the Lemaître–Tolman–Bondi (LTB) models for inhomogeneous matter distribution. This approach was generally explored in Krasinski & Hellaby (2004) Plebanski & Krasinski (2006), and Jacewicz & Krasinski (2012) and seems to be rather promising.

Following the essence of this approach, in the present paper we derive and study the exact solution of the LTB class for a black hole in nonstatic dust cosmological background. In our model, we suppose the black hole to be Schwarzschild-like, positioned in the center of symmetry of the homogeneously distributed cosmological dust. Although the dust background is supposed to be homogeneous, an inhomogeneity occurs when one puts the black hole in the center, and hence the whole model becomes inhomogeneous. The background dust is distributed in such a way that the average distance between the dust particles and the distance from the black hole to the nearest dust particle are much bigger than the gravitational radius of the black hole (see Figure 1). Thus the dust particles do not interact with the black hole directly, but contribute to the universe expansion. Despite of seeming discontinuity the continuous limit holds valid due to cosmological scales of the model.

The distinguishing point of our analysis is that we propose an idea of how to manage the arbitrariness in the LTB model, which leads to strict and very clear physical interpretation. The method we use to obtain the appropriate solution of the Einstein equations for the system of two sources (black hole and cosmological dust) involves one of fundamental scalar invariants existing for spherically symmetric metrics, usually mentioned as the mass function, which possess the property of additivity in the case of negligible interaction. This method will be considered in more detail in Section 2 of this paper where the exact solution for our model will be derived.

The paper is organized as follows. In Section 3, we discuss the properties of the obtained solution. It is shown that near the center of symmetry the solution tends to the Schwarzschild one, while at distant regions the energy density of the dust is Friedmannian. We also built the R–T structure of the resulting spacetime and demonstrate the behavior of the horizons. The geodesic equations for a test particle in comoving coordinates are found and investigated for the three types of spatial curvature in Section 4. The motion of the test particles near the black hole from the point of view of the distant observer and correspondent trajectories is considered in Section 5. Finally, in Section 6, we summarize the results.

## 2. Mass Function Method

Let us first consider a spherically symmetric interval of the general form

$$ds^2 = e^{\nu(R,t)} dt^2 - e^{\lambda(R,t)} dR^2 - r^2(R, t) d\sigma^2, \quad (1)$$

where $d\sigma^2 = d\theta^2 + \sin^2\theta d\varphi^2$ is the standard metric on a unit 2-sphere. It was shown in Cahill & McVittie (1970) that for the interval (1) there exists a combination $rR^{\varphi}_{\theta\varphi\theta}$ that is invariant relative to the coordinate transformations of the form $\bar{t} = \bar{t}(t, R)$, $\bar{R} = \bar{R}(t, R)$, where $R^{\varphi}_{\theta\varphi\theta}$ is a single component of the Riemann–Christoffel tensor. Taking into account the expression for $R^{\varphi}_{\theta\varphi\theta}$, it is possible to introduce the so-called mass function (also known as the Misner–Sharp mass; Misner & Sharp 1964; Zannias 1990) which is invariant provided the spherical symmetry is retained

$$m(R, t) = r(R, t)(1 + e^{-\nu(R,t)} \dot{r}^2 - e^{-\lambda(R,t)} r'^2). \quad (2)$$

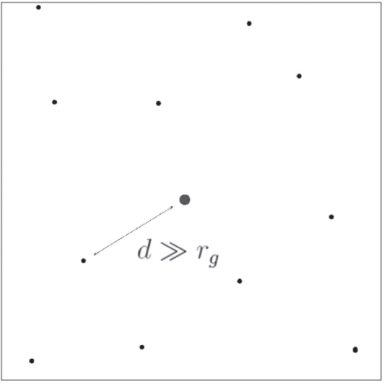

**Figure 1.** Schwarzschild-like black hole immersed in the homogeneous cosmological dust background. The average distance $d$ between the dust particles and the distance from the nearest dust particle to the black hole are much bigger than the gravitational radius of the black hole.

This equation was originally developed in Hernandez & Misner (1966). Throughout this paper, dot and prime refer to the partial derivatives with respect to $t$ and $R$, respectively, and the system of units is used where $c = 8\pi G = 1$.

In terms of the mass function (2), the Einstein equations may be rewritten in a much simpler way

$$m' = \varepsilon\, r^2\, r', \quad (3)$$

$$\dot{m} = -p_\parallel\, r^2\, \dot{r}, \quad (4)$$

$$2\dot{r}' = \nu'\, \dot{r} + \dot{\lambda}\, r', \quad (5)$$

$$2\dot{m}' = m' \frac{\dot{r}}{r'} \nu' + \dot{m} \frac{r'}{\dot{r}} \dot{\lambda} - 4r\, \dot{r}\, r'\, p_\perp, \quad (6)$$

where $\varepsilon$ is the energy density, $p_\perp$ is the tangential pressure and $p_\parallel$ is the radial pressure. Regarding the Equation (3), the mass function is interpreted as the total mass (including the gravitational energy) enclosed in the shell $R = \mathrm{const}$:

$$m = \int_0^R \varepsilon\, r^2\, r' dR. \quad (7)$$

From here, in particular, it follows that in the case of negligible interaction the mass function possesses the property of additivity. Indeed, if one considers the system of several sources that do not interact with each other, then the total energy density will be the sum of the energy densities of the sources. Constructing the general solution for several sources, one should form the correspondent system of Equations (3)–(6) with summarized energy density. As far as the mass function is invariant with respect to $\{R, t\}$ coordinate transformations, from (3) it will follow that the mass function for the whole system is the sum of mass functions of each source.

The model of our interest contains two sources: the central black hole and the cosmological dust surrounding. To describe such a model, we chose the solution of the LTB class (Tolman 1969) for the inhomogeneous dust distribution. The general LTB metric written in comoving coordinates reads

$$ds^2 = dt^2 - \frac{r'^2(R, t)}{f^2(R)} dR^2 - r^2(R, t)\, d\sigma^2, \quad (8)$$

were $f(R)$ is an arbitrary function having the sense of the total energy (in dimensionless units) of the particle in the shell $R = \mathrm{const}$. Depending on the sign of the expression $f^2 - 1$, the metric (8) represents a spacetime with zero ($f^2 = 1$), positive ($f^2 < 1$) or negative ($f^2 > 1$) curvature of the spatial part.

According to the definition (2) the mass function in this case will have the form

$$m(R) = r(R, t)[\dot{r}^2(R, t) + 1 - f^2(R)]. \quad (9)$$

As far as the dust matter is characterized by zero pressure, the mass function (9) will only depend on the $R$ coordinate (see Equation (4)). Equation (9) may be easily integrated, producing three types of the Tolman solution depending on the sign of $f^2 - 1$:

1. The parabolic type

$$f^2(R) = 1,$$
$$r(R, t) = \left[\pm \frac{3}{2} \sqrt{m(R)}\, (t - t_0(R))\right]^{\frac{2}{3}}. \quad (10)$$

2. The elliptic type

$$f^2(R) < 1,$$
$$r(R, t) = \frac{m(R)}{1 - f^2(R)} \sin^2 \frac{\alpha}{2},$$
$$t - t_0(R) = \pm \frac{m(R)}{2(1 - f^2(R))^{3/2}} (\alpha - \sin\alpha). \quad (11)$$

3. The hyperbolic type

$$f^2(R) > 1,$$
$$r(R, t) = \frac{m(R)}{f^2(R) - 1} \sinh^2 \frac{\alpha}{2},$$
$$t - t_0(R) = \pm \frac{m(R)}{2(f^2(R) - 1)^{3/2}} (\sinh\alpha - \alpha). \quad (12)$$

Here, $m(R)$, $f(R)$, and $t_0(R)$ appear as arbitrary functions of integration and the signs $\pm$ correspond to the expanding/contracting universe, respectively.

As is known (see, e.g., Stephani et al. 2009, Section 15.5), the Tolman solution contains the following important particular cases:

1. the Schwarzschild solution corresponding to the choice $m(R) = r_g$, where $r_g$ is the Schwarzschild radius;
2. the Friedmann solution under the following choice of the arbitrary functions: $t_0(R) = 0$, in addition to $f(R) = 1$ and $m(R) = a_0 R^3$ for the parabolic type, $f(R) = \cos R$ and $m(R) = a_0 \sin^3 R$ for the elliptic type, and finally $f(R) = \cosh R$ and $m(R) = a_0 \sinh^3 R$ for the hyperbolic type, where $a_0$ is a constant that is associated with the current size of the universe in Friedmann models.

Now regarding the additivity of the mass function in our case, we may obtain a solution for the dust-filled universe with the immersed black hole substituting into (10)–(12) the combination of mass functions $m_{\text{Schwarzschild}} + m_{\text{Friedmann}}$.

Thus we shall make the following choice of arbitrary functions $m(R)$ and $f(R)$ for each type of spatial curvature

$$m(R) = r_g + a_0 \begin{cases} R^3, & f(R) = 1, \\ \sin^3 R, & f(R) = \cos R, \\ \sinh^3 R, & f(R) = \cosh R. \end{cases} \quad (13)$$

These expressions, together with (10)–(12), give us three types of the desired solution, which was originally obtained in Korkina & Kopteva (2012) and studied for the flat case in Jaluvkova et al. (2017).

## 3. Properties of Solution and R–T Structure of Spacetime

We shall now show that the obtained solution indeed describes a Schwarzschild-like black hole immersed into the Friedmann background. Substituting (13) into (10)–(12), for an expanding universe with the interval (8), we obtain:

1. For a flat space:

$$f(R) = 1,$$
$$m(R) = r_g + a_0 R^3,$$
$$r(R, t) = \left[ \frac{3}{2} \sqrt{r_g + a_0 R^3} \, (t - t_0(R)) \right]^{\frac{2}{3}}. \quad (14)$$

2. For a space with positive curvature (closed space):

$$f(R) = \cos R, \qquad m(R) = r_g + a_0 \sin^3 R,$$
$$r(R, t) = \frac{r_g + a_0 \sin^3 R}{\sin^2 R} \sin^2 \frac{\alpha}{2},$$
$$t - t_0(R) = \frac{r_g + a_0 \sin^3 R}{2 \sin^3 R} (\alpha - \sin \alpha). \quad (15)$$

3. For a space with negative curvature (open space):

$$f(R) = \cosh R, \qquad m(R) = r_g + a_0 \sinh^3 R,$$
$$r(R, t) = \frac{r_g + a_0 \sinh^3 R}{\sinh^2 R} \sinh^2 \frac{\alpha}{2},$$
$$t - t_0(R) = \frac{r_g + a_0 \sinh^3 R}{2 \sinh^3 R} (\sinh \alpha - \alpha). \quad (16)$$

Using a series expansion, it is easy to see that all solutions (14)–(16) exhibit (within a second-order accuracy) the same limit near the center $R \ll 1$:

$$t - t_0(R) = \frac{2 r^{3/2}}{3 \sqrt{r_g}} + k R^2 \frac{r^{5/2}}{5 r_g^{3/2}}, \quad (17)$$

where $k$ is the curvature factor appearing near small corrections for the curved spaces: $k = 0$ for a flat space, $k = 1$ for a closed space, and $k = -1$ for an open space. The first term in (17) is nothing but the parabolic type of the Schwarzschild solution in comoving coordinates, which gives the Lemaître solution (Lemaître 1933) under $t_0(R) = R$. The second term is a correction that will also arise if we consider the elliptic and hyperbolic types of the Schwarzschild solution in comoving coordinates at small $R$.

Thus, by the very natural procedure of calculating the asymptotic it is shown that the solution (14)–(16) near the center of symmetry turns out to be very close to the Schwarzschild one. One should mention here that the arbitrary functions $t_0(R)$ are supposed to be chosen in a particular way to avoid the problems with singularities and to lead to the same expression in the limit $R \to 0$. The possibility of such a choice will be discussed further.

For any shell $R = \text{const}$, a sufficiently large distance $r(R, t)$ corresponds to a sufficiently long time. Let us estimate the energy density of the background dust at distant regions. From Equation (3), it follows that

$$\varepsilon = \frac{m'}{r^2 \, r'}. \quad (18)$$

Assuming $a_0 \gg r_g$, which is the case for the real universe, we obtain the energy density for the three types of spatial curvature in the form:

$$\varepsilon = \frac{4 a_0 R^2}{(t - t_0(R))[3 a_0 R^2 (t - t_0(R)) - 2(a_0 R^3 + r_g) t_0'(R)]}, \quad k = 0, \quad (19)$$

$$\varepsilon = \frac{4 \cosh R}{(t - t_0(R))[3 \cosh(R)(t - t_0(R)) - 2 \sinh(R) t_0'(R)]}, \quad k = -1, \quad (20)$$

$$\varepsilon = \frac{4 \cos R}{(t - t_0(R))[3 \cos(R)(t - t_0(R)) - 2 \sin(R) t_0'(R)]}, \quad k = 1. \quad (21)$$

Expanding into series, the limit of long time gives the same result for each expression (19)–(21):

$$\varepsilon = \frac{4}{3 t^2} + \mathcal{O}\left( \frac{1}{t^3} \right). \quad (22)$$

The first term here is the energy density of a homogeneous dust in the Friedmann model.

We conclude that the obtained inhomogeneous solution (14)–(16) of LTB class indeed describes the Schwarzschild-like black hole in the center of symmetry and homogeneous dust background in distant regions. We stress that, unlike the McVittie's solution and its modifications, we have the surroundings filled with pure dust with zero pressure; it can be seen from Equation (4) with the mass function (13).

Being a particular case of the Tolman solution, the solution (14)–(16) has two true singularities. The Big Bang singularity $r(R, t) = 0$ occurs for each shell $R = \text{const}$ at the time $t = t_0(R)$. The shell-crossing singularity $r'(R, t) = 0$ occurs in the process of universe evolution when the neighboring shells overtake each other producing a singular surface where the energy density, pressure, and spatial curvature exhibit a pathological behavior. Avoiding the shell-crossing in LTB models was widely discussed in the literature (see, e.g., Hellaby & Lake 1985; Nolan 2003). We comment on this issue separately for each type of solution.

For the LTB solutions there also exists a coordinate condition that indicates the boundary between the R and T regions of the spacetime (Novikov 2001), which is equivalent to the event horizon and appears as a removable singularity of the metric when one switches to the curvature coordinates. In any other coordinate the existence of such regions is reflected by the fact that in the T region, which is the region of essential nonstationarity, a static observer is impossible. The boundary between the R and T regions is defined by the "horizon equation"

$$m(R) = r(R, t), \quad (23)$$

which, for example, for the Schwarzschild solution yields the known expression $r = r_g$.

Hence, for the metric (8) there exist three special time instants: the moment of passing through the central singularity $t_0(R)$, the moment of passing through the shell-crossing singularity $t_{sc}(R)$, and the moment of passing through the horizon $t_{hor}(R)$, where $t_{sc}(R)$ is a solution of the equation

$$r'(R, t) = 0 \quad (24)$$

and $t_{hor}(R)$ is a solution of the "horizon equation" (23). In order to avoid the shell-crossing and to satisfy the cosmic censorship conjecture, one could require

$$t_{sc}(R) < t_0(R) < t_{hor}(R). \quad (25)$$

The first inequality here means that at any $R$ the shell-crossing "happens earlier" than the universe was born, i.e., the line of shell-crossings moves to the nonphysical region. The second inequality naturally means that the initial singularity for each shell is hidden under the horizon for any sufficiently distant observer.

The requirement (25) may be treated as a condition for the appropriate choice of the arbitrary function $t_0(R)$. Let us now find the general expressions for $t_{sc}(R)$ and $t_{hor}(R)$ for each type of solution (14)–(16). For further purposes, it is convenient to rewrite the solution in dimensionless units

$$\tilde{r} = \frac{r}{r_g}, \quad \tilde{t} = \frac{t}{r_g}, \quad \tilde{m} = \frac{m}{r_g}, \quad \beta = \frac{a_0}{r_g}. \quad (26)$$

In the rest of our paper, we shall use these dimensionless units and thus prefer to omit the tilde over the letters for brevity. The solution (14)–(16) in dimensionless units (26) reads:

1. The case of flat space:

$$m(R) = 1 + \beta R^3,$$
$$r(R, t) = \left[\frac{3}{2}\sqrt{1 + \beta R^3}\,(t - t_0(R))\right]^{\frac{2}{3}}. \quad (27)$$

2. The case of closed space:

$$m(R) = 1 + \beta \sin^3 R,$$
$$r(R, t) = \frac{1 + \beta \sin^3 R}{\sin^2 R} \sin^2 \frac{\alpha}{2},$$
$$t - t_0(R) = \frac{1 + \beta \sin^3 R}{2 \sin^3 R}(\alpha - \sin \alpha). \quad (28)$$

3. The case of open space:

$$m(R) = 1 + \beta \sinh^3 R,$$
$$r(R, t) = \frac{1 + \beta \sinh^3 R}{\sinh^2 R}\sinh^2 \frac{\alpha}{2},$$
$$t - t_0(R) = \frac{1 + \beta \sinh^3 R}{2 \sinh^3 R}(\sinh \alpha - \alpha). \quad (29)$$

By solving Equations (23) and (24) successively for each case (27)–(29), we obtain:

1. For $k = 0$

$$t_{hor}(R) = t_0(R) + \frac{2}{3}(1 + \beta R^3),$$
$$t_{sc}(R) = t_0(R) + \frac{2}{3}\left(\frac{1}{\beta R^2} + R\right)t_0'(R). \quad (30)$$

2. For $k = 1$

$$t_{hor1}(R) = t_0(R) + \beta R + R\csc^3 R$$
$$\quad - \beta \sin R \cos R - \cot R \csc R,$$
$$t_{hor2}(R) = t_0(R) + (\pi - R)(\beta + \csc^3 R)$$
$$\quad + \beta \sin R \cos R + \cot R \csc R,$$
$$t_{sc1}(R) = t_0(R) + \frac{1}{15\beta}[10(\beta \tan R + \csc^2 R \sec R)$$
$$\quad + 3\beta^{1/3}t_0'(R)^{2/3}\tan^{5/3}R]t_0'(R),$$
$$t_{sc2}(R) = t_0(R) + \pi(\beta + \csc^3 R)$$
$$\quad - \frac{1}{15\beta}[10(\beta \tan R + \csc^2 R \sec R)$$
$$\quad + 3\beta^{1/3}t_0'(R)^{2/3}\tan^{5/3}R]t_0'(R). \quad (31)$$

3. For $k = -1$

$$t_{hor}(R) = t_0(R) + \frac{1 + \beta \sinh^3 R}{2 \sinh^3 R}(\sinh 2R - 2R),$$
$$t_{sc}(R) = t_0(R) - \frac{1}{5\beta^{2/3}}t_0'^{5/3}(R)\tanh^{5/3}R$$
$$\quad + \frac{2}{3}t_0'(R)\tanh R. \quad (32)$$

Here, in our calculations, we have used the fact that $\beta$ is very large for realistic objects (estimative size of the universe is about $a_0 \sim 10^{26}$ m, and $r_g$ of a typical black hole in the galaxy center is about $\sim 10^{10}$ m, i.e., one has $\beta \sim 10^{16}$). Thus, we have employed the Taylor expansion for large $\beta$ to get rid of transcendental equations for the open and closed spaces.

Analyzing the expressions (30)–(32) and applying the condition (25), we come to the following form of the arbitrary functions $t_0(R)$:

$$t_0(R) = \frac{1}{1 + R^{3/2}}, \quad k = 0, \quad (33)$$

$$t_0(R) = \frac{1}{1 + \sin^{3/2} R}, \quad k = 1, \quad (34)$$

$$t_0(R) = \frac{1}{1 + \sinh^{3/2} R}, \quad k = -1. \quad (35)$$

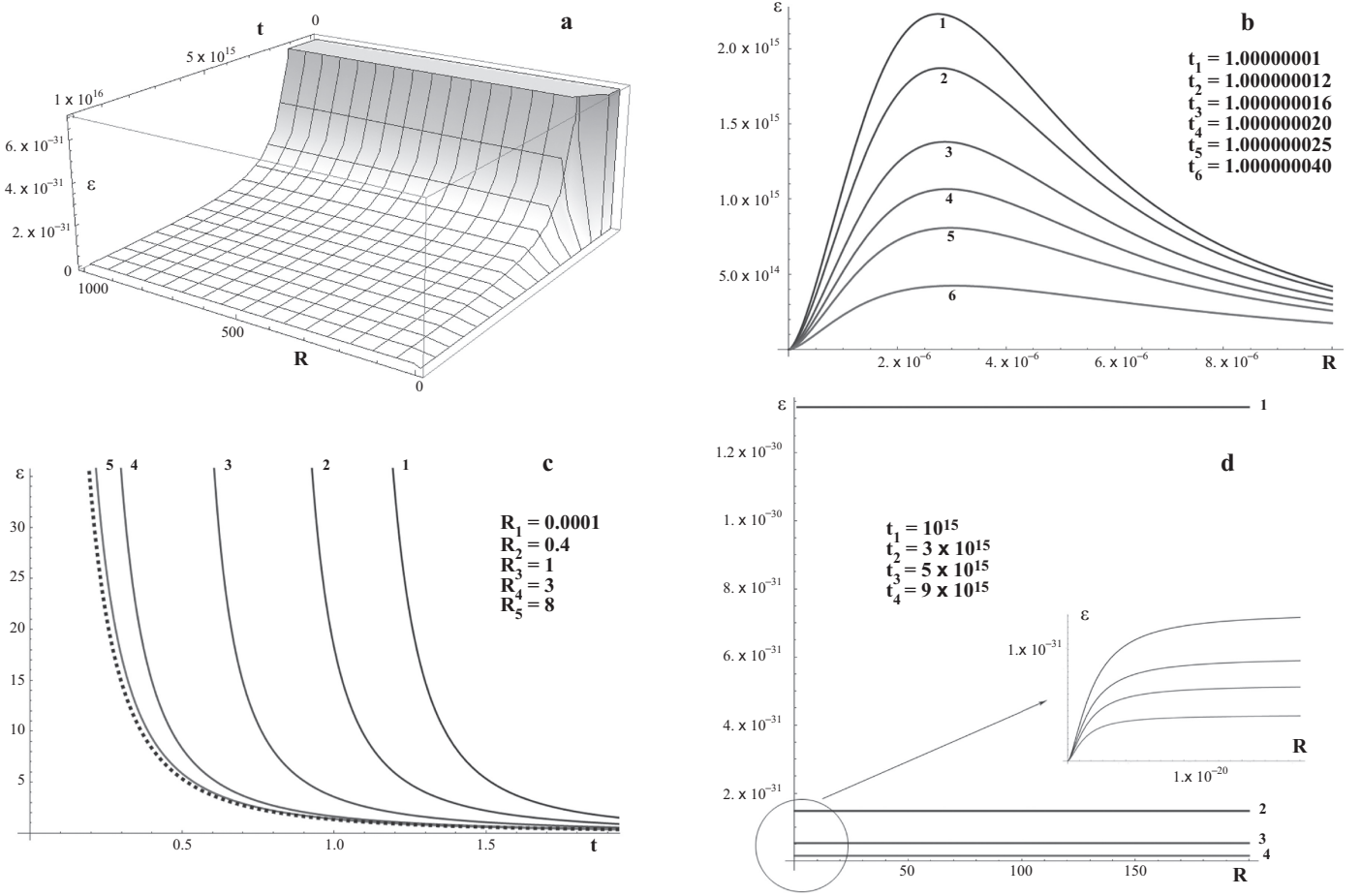

**Figure 2.** Energy density profile for the parabolic case of our solution given in dimensionless units. (a) At the initial (Big Bang) singularity there is an infinite value of the energy density that is decreasing with time, while in the center of symmetry there always holds $\varepsilon(R = 0, t) = 0$. (b) The structure of the profile at the initial singularity in the vicinity of the central point $\{R = 0, t = 1\}$. There is a punctured infinite energy density cusp, which is rapidly decreasing with time. (c) The character of the energy density dependence on time for different values of $R$. The dashed curve corresponds to the parabolic Friedmann solution. The bigger is $R$, the closer is $\varepsilon(R, t)$ to the Friedmannian line. (d) The picture at the epoch of the Milky Way birth. The number of each curve corresponds to the index of the value $t$ (or $R$) listed in the legend.

It is easy to see that these functions have the same limit at small $R$ and the result (17) remains valid.

To give an example of the behavior of the obtained model, we present in Figures 2–3 the profiles of the energy density and the Hubble parameter for the parabolic case of our solution (14). The energy density is given by (19). The Hubble parameter describing the rate of expansion of the inhomogeneous universe is generally defined according to Ellis (2009) as follows

$$H = \frac{1}{l}\frac{dl}{d\tau} = \frac{1}{3}\Theta, \qquad (36)$$

where $l$ is some characteristic length that corresponds to the scale factor $a(t)$ in Friedmann models, and $\tau$ is proper time given in the standard way:

$$d\tau = \sqrt{g_{00}}\,dt. \qquad (37)$$

$\Theta$ is the volume expansion given by

$$\Theta = u^\alpha_{;\alpha}, \qquad (38)$$

where $u^\alpha$ is usual 4-velocity of the dust particles.

In our case for the metric (8) with coefficients found from (14), we obtain

$$H = \frac{1}{2\sqrt{g_{00}}}\frac{1}{g_{11}}\frac{\partial}{\partial t}g_{11}$$
$$= \frac{(R^{3/2} + 1)(\beta R^{3/2}(R^{3/2}(4t - 3) + 2R^3 t + 2(t - 1)) - 1)}{3(R^{3/2}t + t - 1)(\beta R^{3/2}((R^{3/2} + 1)^2 t - 1) + 1)} \qquad (39)$$

We would like to mention here that the profiles presented in Figures 2–3 should be treated as "possible for observations" beginning from the appropriate timescale when the ordinary black holes were formed. The specific initial features were demonstrated to give the more complete information about the model and its self-consistency.

It is seen (Figure 2(d)) that there is a region of low density around a black hole (analogous to the Schwarzschild vacuole in the ES model) that appears naturally in our case.

In order to illustrate the structure of our model spacetime, we also present two-dimensional sections in the $(R, t)$ plane where the R–T regions with horizons and singularities are built. In Figures 4–5, the R–T regions for the ordinary Schwarzschild and Friedmann solutions (the parabolic types) are shown just

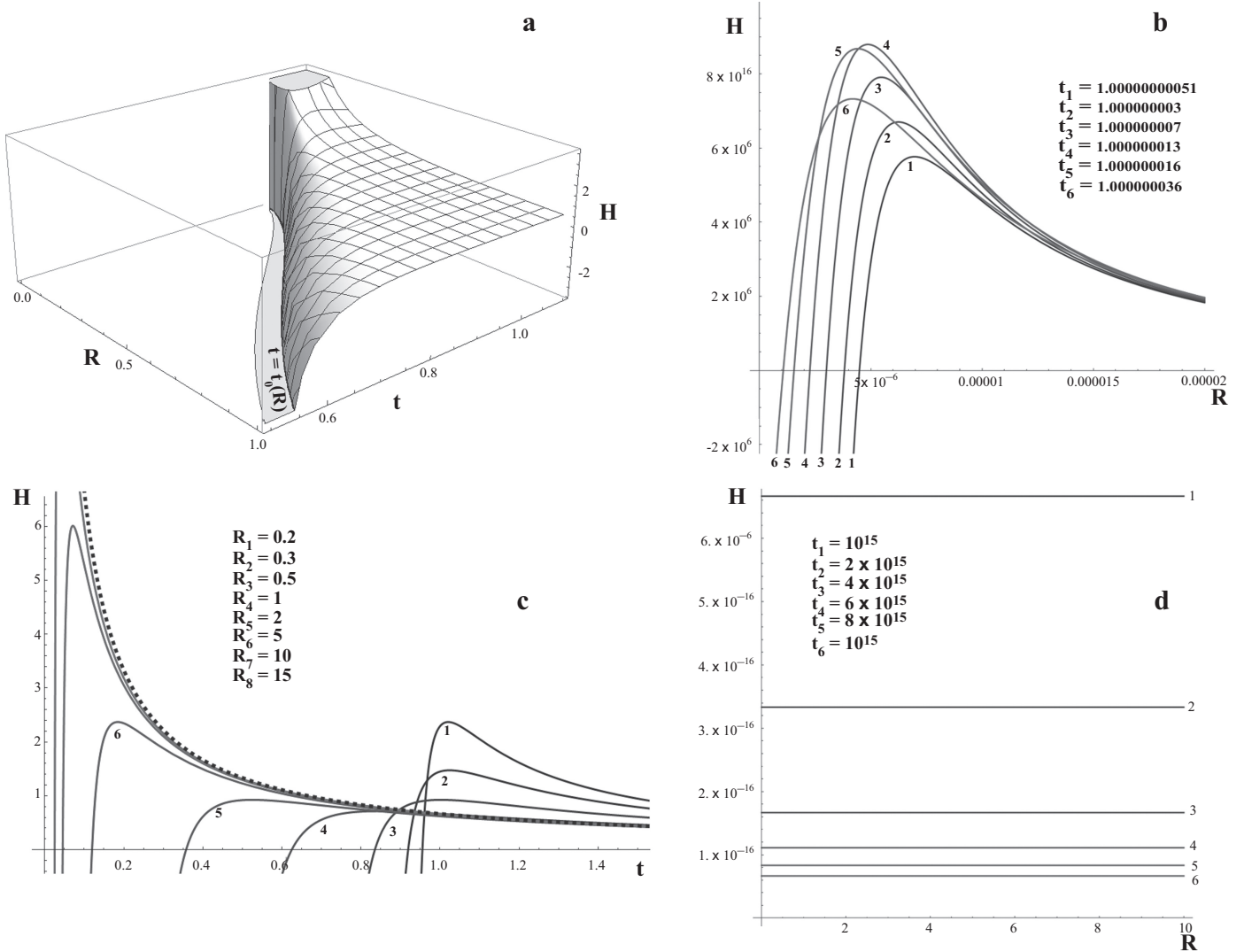

**Figure 3.** The Hubble parameter profile for the parabolic case of our solution given in dimensionless units. (a) The line $t = t+0(R)$ marks the line of the initial singularity. (b) The structure of the profile at the initial singularity in the vicinity of the central point $\{R = 0, t = 1\}$. (c) The Hubble parameter dependence on time for different values of $R$. The dashed curve corresponds to the Hubble parameter of the parabolic Friedmann solution. The bigger is $R$, the closer is $H(R, t)$ to the Friedmannian curve. (d) The picture at the epoch of the Milky Way birth. The number of each curve corresponds to the index of the value $t$ (or $R$) listed in the legend.

for comparison with the "combined" spacetime of the solution (14)–(16) (Figures 6–8).

In Figure 4, the arbitrary function $t_0(R)$ in the Scwarzschild solution is chosen in the same way as (33) and the sign is chosen for the case of expansion. The event horizon is parallel to all the other lines of constant $r$ and obviously to the line of singularity $r = 0$. The rate of expansion is the same for all shells $R = $ const. The shell-crossing is absent for this solution. The similar picture is often occurring for the Lemaître solution with $t_0(R) = R$ when the process of gravitational collapse is considered (Landau & Lifshitz 1975, Section 102; Misner et al. 1973, Section 31.4).

Figure 5 shows the R–T structure of the Friedmann spacetime in the parabolic case. The universe starts simultaneously at all shells $R = $ const but then the rate of expansion changes for different shells. The horizon runs away from the initial singularity as $t \sim R^3$.

Figure 6 shows the R–T structure of the spacetime of the parabolic type of our solution (14). It is seen that the behavior of the horizon (the line $r = m$) in the central region ($R \ll 1$, the shadowed part of the diagram) is similar to the Schwarzschild picture, while in distant regions the Friedmannian behavior prevails. The universe in each shell $R = $ const starts at a different time $t_0(R)$. The line of shell-crossings $r' = 0$ lies below the line of the initial singularity $r = 0$, i.e., belongs to the nonphysical region.

In Figure 7, the same is depicted for the elliptic type of our solution (15). The situation here is more complicated as far as the universe in this case should recollapse. Thus there will be two singularities: the initial and the final one, and hence two horizons appear dividing the spacetime into two R and two T regions. The similar picture for the elliptic case of the Schwarzschild solution was analyzed in Novikov (1963) in comoving coordinates for the observer moving along the bounded radial trajectory. Figure 7(b) demonstrates the behavior of the horizon with respect to the initial singularity in the central region ($R \ll 1$). Figures 7(c), (d) confirm the nonphysical positions of shell-crossings relating to the initial (c) and the final (d) singularities.

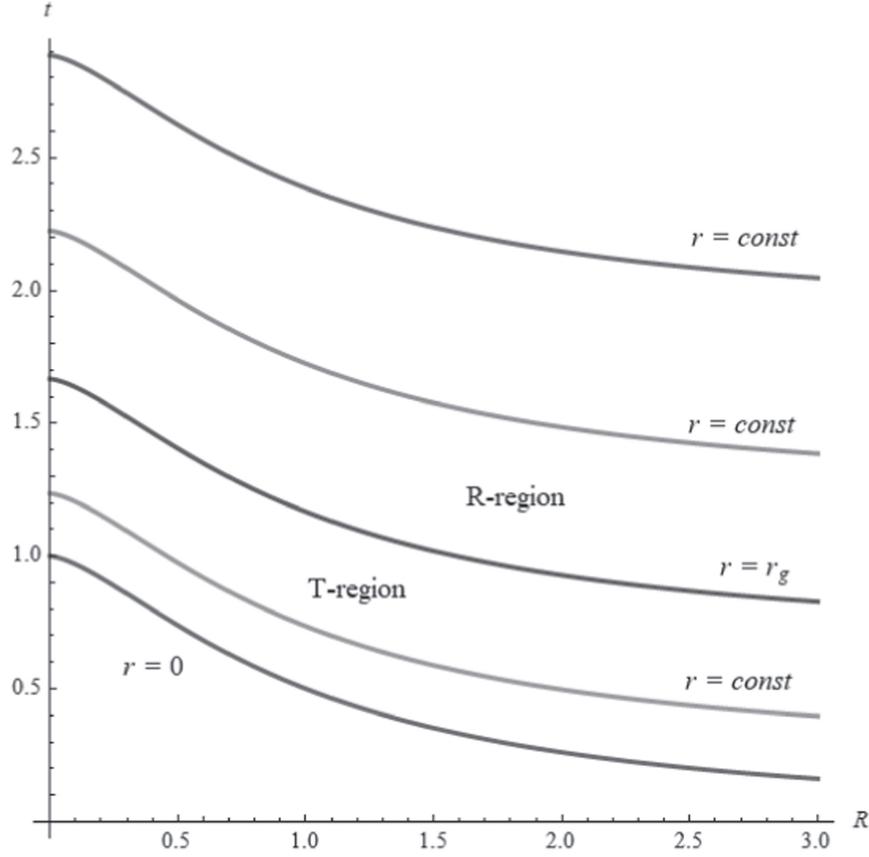

**Figure 4.** R–T regions for Schwarzschild parabolic solution in dimensionless units. Horizon is parallel to lines of constant "radius." The rate of expansion is the same for all shells. Shell-crossing singularity is absent in this solution.

Figure 8 represents the same for the hyperbolic type of our solution (16).

It can be seen from these illustrations that for any observer from the R region there will always be a region in the center where the spacetime is Schwarzschild-like. And if the observer is distant enough, he will see the usual behavior of the test particles as near the Schwarzschild event horizon.

### 4. Motion of Test Particles: Comoving Observer

In this section, we shall derive the equations of motion of a test particle in our model from the point of view of a comoving observer, i.e., a local observer at rest relative to the averaged motion of the neighbor matter. We fix the $\theta$ coordinate as $\theta = \pi/2$, which corresponds to a motion in the equatorial plane, and choose the arbitrary function $t_0(R)$ according to (33)–(35).

For the solution (27)–(29) with the interval (8) the geodesic equations found by the standard method read

$$\frac{d^2 t}{ds^2} = -\frac{\dot r' r'}{f^2}\left(\frac{dR}{ds}\right)^2 - \dot r r \left(\frac{d\varphi}{ds}\right)^2, \tag{40}$$

$$\frac{d^2 R}{ds^2} = \left(\frac{f'}{f} - \frac{r''}{r'}\right)\left(\frac{dR}{ds}\right)^2 + f^2 \frac{r}{r'}\left(\frac{d\varphi}{ds}\right)^2$$
$$- 2 \frac{\dot r'}{r'} \frac{dt}{ds}\frac{dR}{ds}, \tag{41}$$

$$\frac{d^2 \varphi}{ds^2} = -2 \frac{r'}{r} \frac{dR}{ds}\frac{d\varphi}{ds} - 2 \frac{\dot r}{r} \frac{d\varphi}{ds}\frac{dt}{ds}. \tag{42}$$

If the particle is at rest with respect to the comoving coordinates $R$ and $\varphi$ one has

$$\frac{dR}{ds} = 0, \qquad \frac{d\varphi}{ds} = 0, \qquad \frac{dt}{ds} = 1, \tag{43}$$

and hence from the system (40)–(42) it follows that

$$\frac{d^2 t}{ds^2} = 0, \qquad \frac{d^2 R}{ds^2} = 0, \qquad \frac{d^2 \varphi}{ds^2} = 0. \tag{44}$$

This means that being initially at rest the particle will remain at rest and follow the cosmological expansion as all matter on average does.

In order to investigate a more general case of arbitrary initial velocity of the particle in the $\theta = \pi/2$ plane, it is more convenient to use Equations (41) and (42) taking into account the interval (8). Let us first introduce the following notation

$$\frac{dR}{dt} \equiv v, \qquad \frac{d\varphi}{dt} \equiv \omega, \qquad \frac{dt}{ds} \equiv x,$$
$$u_1 \equiv r'v, \qquad u_3 \equiv r\omega, \tag{45}$$

where $u_1$ and $u_3$ are the observable radial and orbital velocities of the test particle, respectively. In terms of the notation (45) Equation (41) may be rewritten in the following form

$$\frac{1}{x}\frac{dx}{dt} = -\frac{1}{v}\frac{dv}{dt} + \left(\frac{1}{f}\frac{df}{dR} - \frac{1}{r'}\frac{\partial r'}{\partial R}\right)v - 2\frac{1}{r'}\frac{\partial r'}{\partial t}$$
$$+ f^2 \frac{r}{r'}\frac{\omega^2}{v}. \tag{46}$$

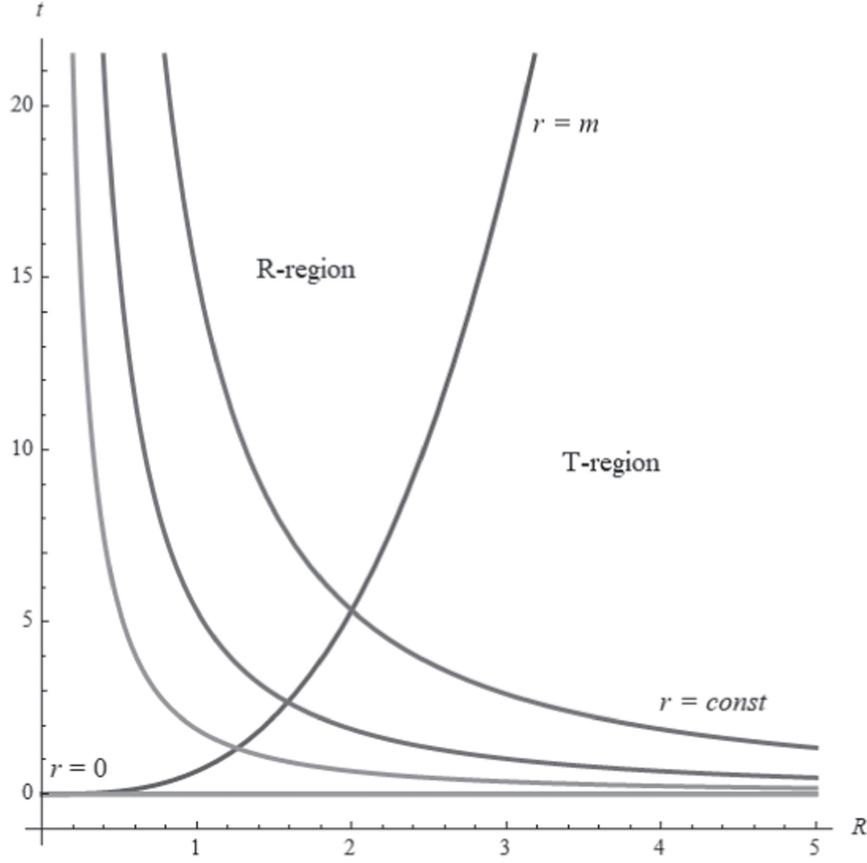

**Figure 5.** R–T regions for Friedmann parabolic solution. Universe starts simultaneously at every shell $R = \text{const}$. The rate of expansion in the center is slower than at distant shells. Shell-crossing singularity is absent in this solution.

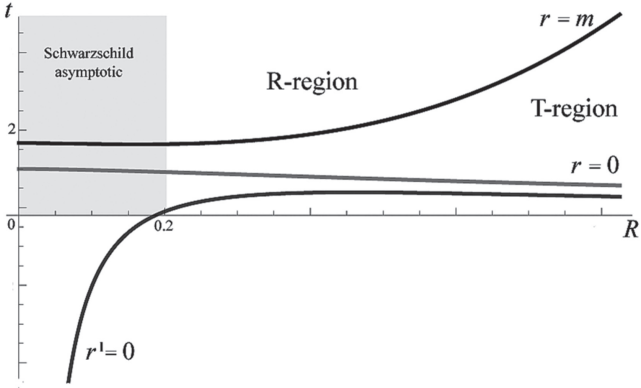

**Figure 6.** R–T structure of spacetime in the model of a cosmological black hole embedded in a dust-filled universe, the flat-space case. The shadowed region corresponds to the Schwarzschild limit of the original solution. The R region is the region of ordinary observers. The shell-crossing singularity "occurs before" the universe is born.

Multiplying (46) by $dt$ gives the equation in differentials

$$\frac{dx}{x} = -\frac{dv}{v} + \frac{df}{f} - \frac{dr'}{r'} - \frac{\partial(\ln r')}{\partial t} dt + f^2 \frac{r}{r'} \frac{\omega}{v} d\varphi,$$

which may be easily integrated with the result

$$\frac{ds}{dt} = C(R)\, r'\, u_1\, e^{-f \int \frac{u_3}{u_1} d\varphi}, \tag{47}$$

where $C(R)$ is an arbitrary function of integration. A similar procedure applied to Equation (42) gives

$$\frac{1}{x}\frac{dx}{dt} = -\frac{1}{\omega}\frac{d\omega}{dt} - \frac{2}{r}\left(\frac{\partial r}{\partial R} v + \frac{\partial r}{\partial t}\right), \tag{48}$$

and after the integration, we obtain

$$\frac{ds}{dt} = A\, r\, u_3, \tag{49}$$

where $A$ is an arbitrary integration constant.

Combining the results of (47) and (49) and taking the logarithm, we obtain

$$-f \int \frac{u_3}{u_1} d\varphi = \ln \frac{Ar}{C(R)\, r'} + \ln \frac{u_3}{u_1}. \tag{50}$$

Taking the derivative with respect to $\varphi$ from both sides of (50), we find the equation that leads to a simple result

$$\frac{u_3}{u_1} = f\varphi + B, \tag{51}$$

with $B$ being an arbitrary constant of integration. From the expression for the interval (8), it follows that

$$\left(\frac{ds}{dt}\right)^2 = 1 - (u_1^2 + u_3^2). \tag{52}$$

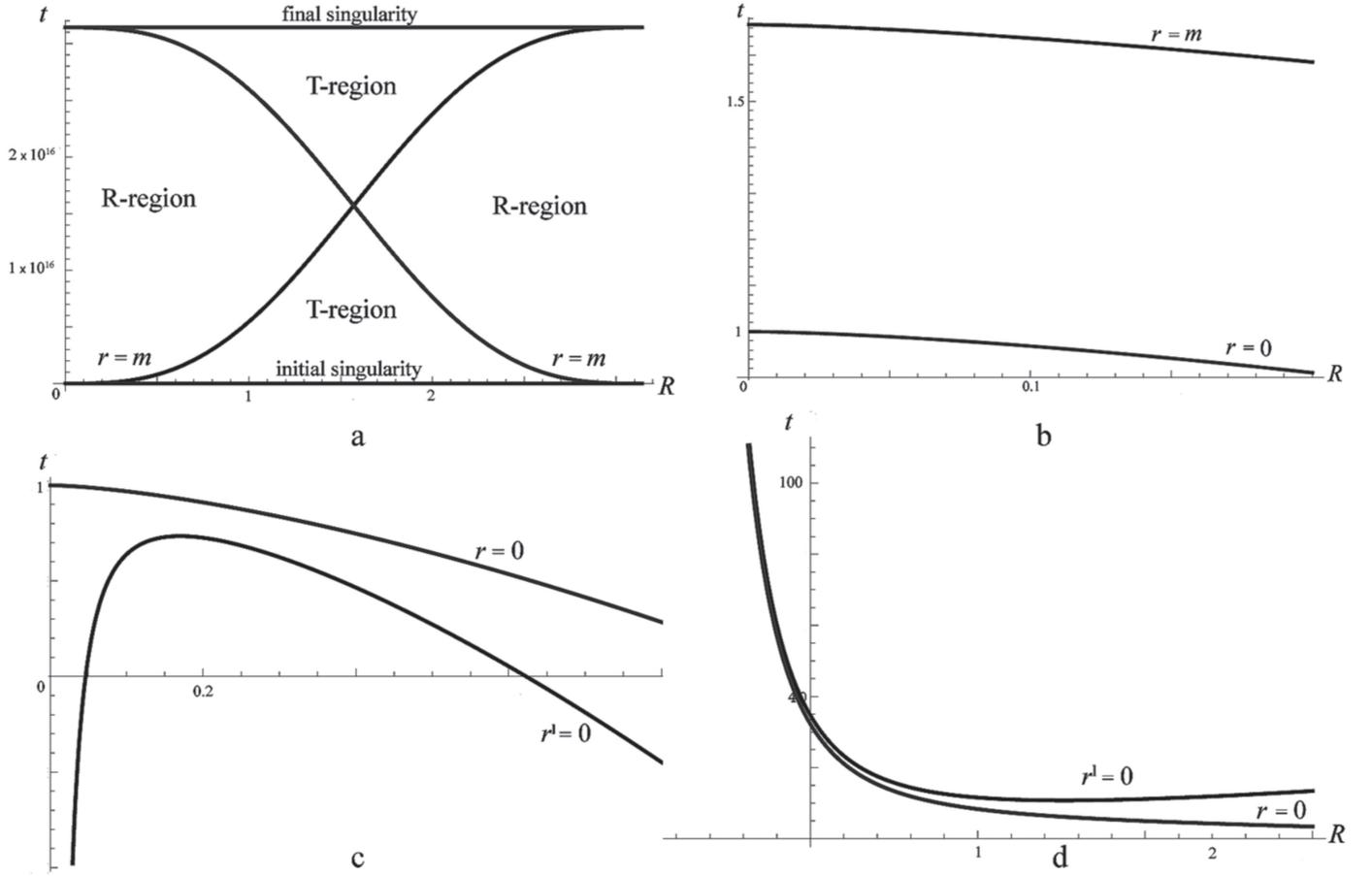

**Figure 7.** Same as Figure 6 but for the closed-space case: (a) R–T structure of the spacetime in the model of a cosmological black hole. (b) Schwarzschild limit of the solution for $R \ll 1$. (c) Initial singularity and shell-crossing. The shell-crossing occurs "earlier" than the Big Bang. (d) Final singularity and shell-crossing. The shell-crossing occurs "later" than the Big Crunch.

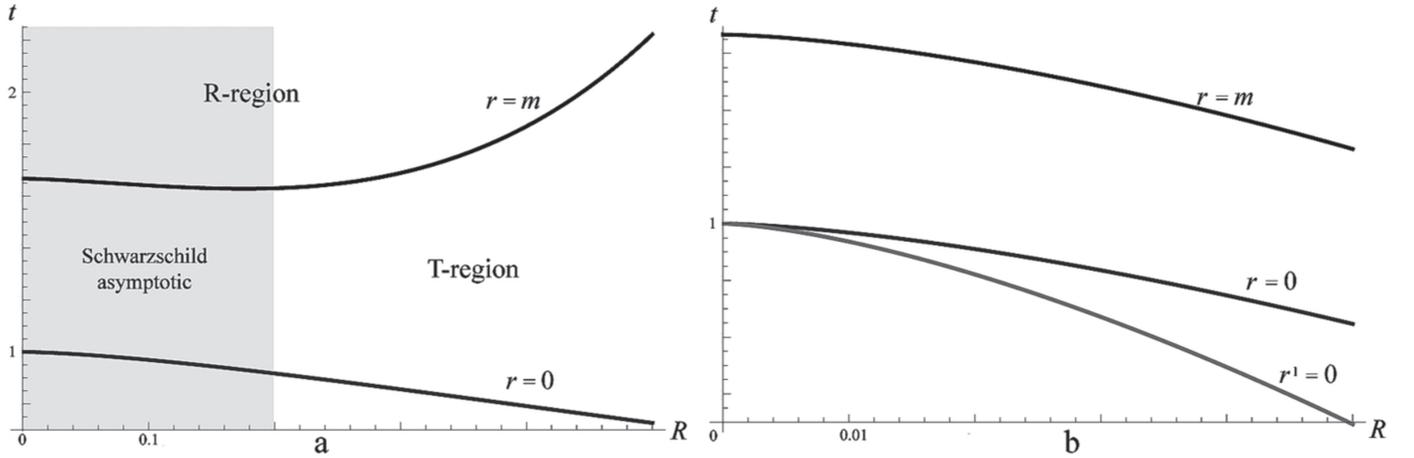

**Figure 8.** Same as Figure 6 but for the open-space case. (a) R–T structure of the spacetime in the model of a cosmological black hole. The shadowed region corresponds to the Schwarzschild asymptotic of the original solution. The R region is the region of ordinary observers. (b) Event horizon, initial singularity and shell-crossing in the Schwarzschild limit. The shell-crossing always "occurs before" the Big Bang.

Combining (49) with (52), we obtain the orbital velocity

$$u_3 = \pm \frac{\sqrt{1 - u_1^2}}{\sqrt{1 + A^2 r^2}}, \qquad (53)$$

and hence the radial velocity becomes

$$u_1 = \pm \frac{f\varphi + B}{\sqrt{1 + A^2 r^2 + (f\varphi + B)^2}}. \qquad (54)$$

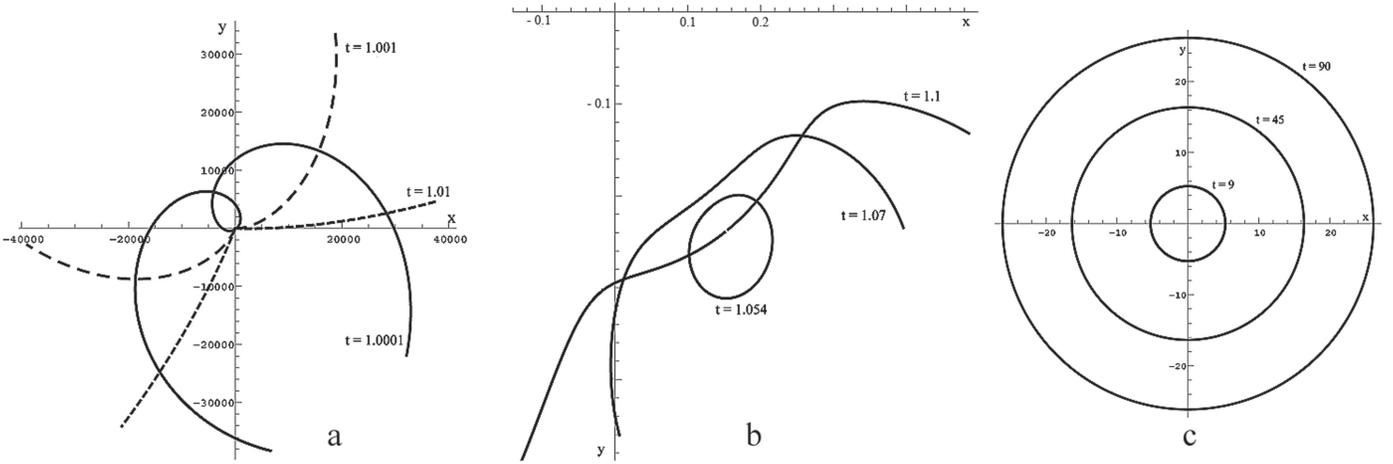

**Figure 9.** Examples of the trajectories of a test particle from the point of view of the comoving observer. $x$ and $y$ are ordinary Cartesian coordinates. (a) The flat-space case: the successive snapshots of the trajectory at the moments of time $t = 1.0001, 1.001, 1.01$ with the initial data: $u_{01} = 0.1$, $u_{03} = 0.1$, $\varphi_0 = 0.001$, $\beta = 10^{16}$. (b) The closed-space case: the successive snapshots of the trajectory at the moments of time $t = 1.054, 1.07, 1.1$ with the initial data: $u_{01} = 0.01$, $u_{03} = 0.01$, $R_0 = 0.0001$, $\varphi_0 = 0.001$, $\beta = 10^{16}$. (c) The open-space case: the successive snapshots of the trajectory at the moments of time $t = 9, 45, 90$ with the initial data: $u_{01} = 0.5$, $u_{03} = 0.5$, $R_0 = 0.0001$, $\varphi_0 = 0.001$, $\beta = 10^{16}$. All values are given in dimensionless units.

Let us assume that at some time $t_{\rm in}$, in which we begin to observe the particle, it occupies the coordinates $R_0$ and $\varphi_0$ with corresponding values of $r(R_0, t_{\rm in}) \equiv r_0$ and $r'(R_0, t_{\rm in}) \equiv r'_0$. Furthermore, let the particle initial velocity and its components be $u_0^2 = u_{01}^2 + u_{03}^2$. An important condition here is $t_{\rm in} > t_0(R_0)$, which means that by the time $t_{\rm in}$ the universe has already started in the shell $R = R_0$. Using these initial conditions, we find from Equations (53) and (54) the values of $A$ and $B$

$$A = \frac{\sqrt{1 - u_0^2}}{r_0 u_{03}}, \qquad B = \frac{u_{01}}{u_{03}} - f_0 \varphi_0. \qquad (55)$$

Finally, for the velocities, one obtains

$$u_1 = \pm \frac{r_0[u_{01} + u_{03}(f\varphi - f_0 \varphi_0)]}{\sqrt{r^2(1 - u_0^2) + r_0^2 \{u_{03}^2 + [u_{01} + u_{03}(f\varphi - f_0 \varphi_0)]^2\}}}, \qquad (56)$$

$$u_3 = \pm \frac{r_0 u_{03}}{\sqrt{r^2(1 - u_0^2) + r_0^2 \{u_{03}^2 + [u_{01} + u_{03}(f\varphi - f_0 \varphi_0)]^2\}}}. \qquad (57)$$

The total observable velocity of the test particle can be found in a standard way to be $u^2 = u_1^2 + u_3^2$. This is the velocity of a free moving particle that would be measured with usual instruments by the set of local observers located along the particle trajectory and being at rest relative to the surrounding medium. The particle trajectories seen by such observers are shown in Figure 9 for the three types of spatial curvature. One should mention here that due to the fact that the metric is nonstatic, the observed trajectory changes with time. Figure 9 shows successive immediate snapshots of the changing trajectories with the correspondent moments of time indicated near each curve.

## 5. Motion of Test Particles: Distant Observer

We now switch from the comoving observer to a reference system of some distant galaxy from which the motion of a test particle near the black hole will be observed. As soon as the galaxy is moving due to the cosmological expansion, the Schwarzschild coordinate system becomes inappropriate. In such a case, the most viable solution is a switch to the curvature coordinates (Gautreau 1984a). For a general nonstatic metric, it is not always possible to make such a coordinate transition in a pure analytical way. A transition was proposed to the resulting metric where only the spatial coordinate is affected, while the global "cosmological" time is preserved (Gautreau 1984b). Although the metric becomes nondiagonal, its advantage is that it involves two coordinates with direct physical interpretation: the curvature coordinate $r$, which is the proper distance for the observer, and time $t$, which continues to be the proper time of the observer. The latter, in particular, allows for a reduction of the order of the geodesic equation for $u^0$ as soon as $dt/dt = 1$. Such a coordinate transition was done, e.g., in Nolan (2014) for the McVittie solution.

Let us make a transition from the metric (8) to a metric of the form

$$ds^2 = g_{00}(r, t)\, dt^2 + 2g_{01}(r, t)\, dr\, dt + g_{11}(r, t)\, dr^2 - r^2\, d\sigma^2. \qquad (58)$$

This can be achieved as follows. First, we take $dr = r'dR + \dot{r}dt$ and expressing $dR$ we substitute it to the metric (8). Thereby, we obtain the expressions for $g_{00}(R, t)$, $g_{11}(R, t)$, and $g_{01}(R, t)$. As a result, for the three types of spatial curvature we find $R(r, t)$ from the corresponding solution (27)–(29) (with respective $t_0(R)$ from (33)–(35)) and substitute it for the expressions of the metric coefficients. Thus, finally, we obtain $g_{00}(R(r, t), t) = g_{00}(r, t)$, $g_{11}(R(r, t), t) = g_{11}(r, t)$, and $g_{01}(R(r, t), t) = g_{01}(r, t)$.

It is easy to check that the metric coefficients in (58) become

$$g_{00} = 1 - \frac{\dot{r}^2}{f^2}, \quad g_{01} = \frac{\dot{r}}{f^2}, \quad g_{11} = -\frac{1}{f^2},$$

$$g_{22} = -r^2, \quad g_{33} = -r^2 \sin^2\theta. \qquad (59)$$

These expressions should be treated somehow formally before the final dependence on $r$ and $t$ is written. The dot here still refers to differentiation with respect to $t$ in the previous metric (8).

As a next step, we need to find $R(r, t)$ from the solutions (27)–(29). For this purpose, we shall make use of two basic assumptions: first, that $\beta$ is very large, and second, that the region of our interest is $R \ll 1$. Keeping the accuracy up to the second order we obtain the following result

$$R(r, t) = \left[\frac{2}{3}r^{3/2} - (t - 1)\right]^{2/3}, \quad k = 0, \quad (60)$$

$$R(r, t) = \frac{\left(\frac{2}{3}\right)^{2/3} r}{\beta^{1/3}(t - 1)^{2/3}}, \quad k = \pm 1. \quad (61)$$

Under the same assumptions, we find $\dot{r}$:

$$\dot{r} = \frac{1}{\sqrt{r}}, \quad k = 0, \quad (62)$$

$$\dot{r} = \sqrt{\frac{1 - r \sin^2 R(r, t) + \beta \sin^3 R(r, t)}{r}}, \quad k = 1, \quad (63)$$

$$\dot{r} = \sqrt{\frac{1 + r \sinh^2 R(r, t) + \beta \sinh^3 R(r, t)}{r}}, \quad k = -1. \quad (64)$$

Finally, the metric coefficients (59) will take the form

1. The case of flat space ($k = 0, f = 1$):

$$\begin{aligned} g_{00} &= 1 - \frac{1}{r}, \\ g_{01} &= \frac{1}{\sqrt{r}}, \\ g_{11} &= -1. \end{aligned} \quad (65)$$

2. The case of closed space ($k = 1, f = \cos R$):

$$\begin{aligned} g_{00} &= 1 - \frac{1}{r} + \left(1 - \frac{1}{r}\right)R^2(r, t), \\ g_{01} &= \frac{1}{\sqrt{r}} - \frac{r - 2}{2\sqrt{r}}R^2(r, t), \\ g_{11} &= -1 - R^2(r, t). \end{aligned} \quad (66)$$

3. The case of open space ($k = -1, f = \cosh R$):

$$\begin{aligned} g_{00} &= 1 - \frac{1}{r} - \left(1 - \frac{1}{r}\right)R^2(r, t), \\ g_{01} &= \frac{1}{\sqrt{r}} + \frac{r - 2}{2\sqrt{r}}R^2(r, t), \\ g_{11} &= -1 + R^2(r, t). \end{aligned} \quad (67)$$

The metric (58) with coefficients (65) after the substitution $t = \tau - 2\left(2\sqrt{r} + \ln\frac{1 - \sqrt{r}}{1 + \sqrt{r}}\right)$ turns into the ordinary Schwarzschild metric in its usual form. Thus, as was expected, in the case of flat space we obtain within a second-order accuracy the same behavior of the test particle near the black hole embedded in a cosmological medium as in the Schwarzschild solution.

To illustrate this fact, we present in Figure 10 the radial coordinate velocity profile $dr/dt$ (in dimensionless units) for the flat-space case of our model

$$\frac{dr}{dt} = \sqrt{1 - \frac{1}{r} - \frac{4r^2}{9(t - 1)^2}} \frac{\sqrt{4r^3 + 9(t - 1)^2}}{3\sqrt{r}(t - 1)}, \quad (68)$$

compared with one built for the ordinary Schwarzschild solution (in case of zero velocity in the infinity)

$$\frac{dr}{dt} = \left(1 - \frac{1}{r}\right)\frac{1}{\sqrt{r}}. \quad (69)$$

The moment of time in Figure 10 for our model is chosen as $t = 10^5$. The difference between profiles is due to the fact that in our model the coordinate $r$ measures the proper radial distance, unlike the Schwarzschild solution, where the proper distance is defined by $\sqrt{g_{11}}dr$. This correction causes both profiles to completely coincide.

For the curved space, it is now possible to write the geodesic equations. Let us again adjust the coordinate system so that the test particle moves in the equatorial plane $\theta = \pi/2$ and hence $u^2 = d\theta/dt = 0$. We should also recount here that in our case it holds that $u^0 = dt/dt = 1$. In terms of $g_{\mu\nu}$ the system of geodesic equations will have the same form for both the closed and open spaces

$$\begin{aligned} \frac{du^0}{dt} &= 0 = \frac{1}{2g_{01}^2 - 2g_{00}g_{11}} \\ &\times [g_{11}(\dot{g}_{00} - (u^1)^2 \dot{g}_{11} + 2u^1 g'_{00} + 2(u^1)^2 g'_{01}) \\ &\quad - g_{01}(2r(u^3)^2 + 2\dot{g}_{01} + 2u^1 \dot{g}_{11} - g'_{00} + (u^1)^2 g'_{11})], \end{aligned} \quad (70)$$

$$\begin{aligned} \frac{du^1}{dt} &= \frac{1}{2g_{01}^2 - 2g_{00}g_{11}} \\ &\times [-g_{01}(\dot{g}_{00} - (u^1)^2 \dot{g}_{11} + 2u^1 g'_{00} + 2(u^1)^2 g'_{01}) \\ &\quad + g_{00}(2r(u^3)^2 + 2\dot{g}_{01} + u^{12} g'_{11})], \end{aligned} \quad (71)$$

$$\frac{du^3}{dt} = -\frac{2u^1 u^3}{r}. \quad (72)$$

Prime and dot refer here, respectively, to the partial derivatives with respect to $r$ and $t$. The Equation (72) can be easily integrated. It leads to the momentum conservation law

$$\frac{du^3}{u^3} = -2\frac{dr}{r},$$

$$u^3 = \frac{L}{r^2}, \quad (73)$$

where $L$ is a constant.

As far as $u^0 = 1$ one can express $u^1$ from Equation (70)

$$\begin{aligned} \frac{dr}{dt} &= \frac{1}{r^3[g_{11}(\dot{g}_{11} - 2g'_{01}) + g_{01}g'_{11}]} \\ &\times [-r^3 g_{01} \dot{g}_{11} + r^3 g_{11} g'_{00} \\ &\quad \pm \frac{1}{2}\{4r^6(g_{01}\dot{g}_{11} - g_{11}g'_{00})^2 \\ &\quad - 4r^3[-r^3 g_{11}\dot{g}_{00} + g_{01}(2L^2 + 2r^3\dot{g}_{01} - r^3 g'_{00})] \\ &\quad \times [g_{11}(\dot{g}_{11} - 2g'_{01}) + g_{01}g'_{11}]\}^{\frac{1}{2}}]. \end{aligned} \quad (74)$$

Equations (73) and (74) together with the metric coefficients (66) or (67) allow us to obtain the trajectories of test particles from the point of view of a distant observer in curved spacetime. The examples of such trajectories are shown in Figures 11 and 12. The set of initial data is represented by momentum $L$, the time of the observations beginning $t_{in}$, initial position of the particle $r(t_{in})$, initial velocity of the particle $r'(t_{in}) \equiv \frac{dr}{dt}|_{t=t_{in}}$. All values are given in dimensionless units (26). Here, we introduce some known parameters expressed in these units for an interested reader to estimate the results. The approximate size of the universe ($\sim 10^{26}$ m) corresponds to $\sim 10^{16}$, the gravitational radius of a typical black hole in the galaxy center ($\sim 10^{10}$ m) equals 1, and the size of the Milky Way ($\sim 10^{21}$ m) reads $\sim 10^{11}$. For the timescale $10^9$ years $\simeq 10^{15}$: the age of the universe $\sim 13 \times 10^9$ years $\simeq 10^{16}$, the age of the Milky Way $\sim 12 \times 10^9$ years $\simeq 10^{16}$, and the age of mankind $\sim 10^5$ years $\simeq 10^{11}$. For example, in Figures 11(a), (b) the particle starts from the distance of about $r_g$ with the initial velocity $\sim 10^{-2}c$ and moves around the elliptic orbit for a long period of time. Figure 11(c) corresponds to the particle starting from a distance of about $2r_g$ with the initial velocity $\sim 10^{-2}c$ and slowly spiralling down to the black hole for a long period of time. The other examples of the trajectories are for the particles with high momentum escaping from the black hole (Figures 11(f) and 12(b)), and for the particles first spiralling down to the black hole and then spiralling away to infinity (Figures 11(d) and 12(a)).

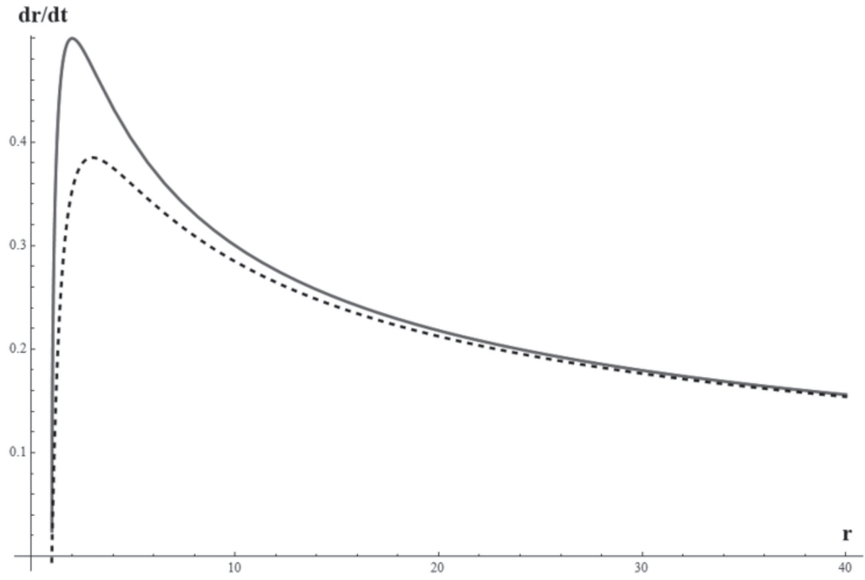

**Figure 10.** Radial velocity profile in the model of a cosmological black hole embedded in a dust-filled universe for the flat-space case. The dashed curve corresponds to the Schwarzschild solution.

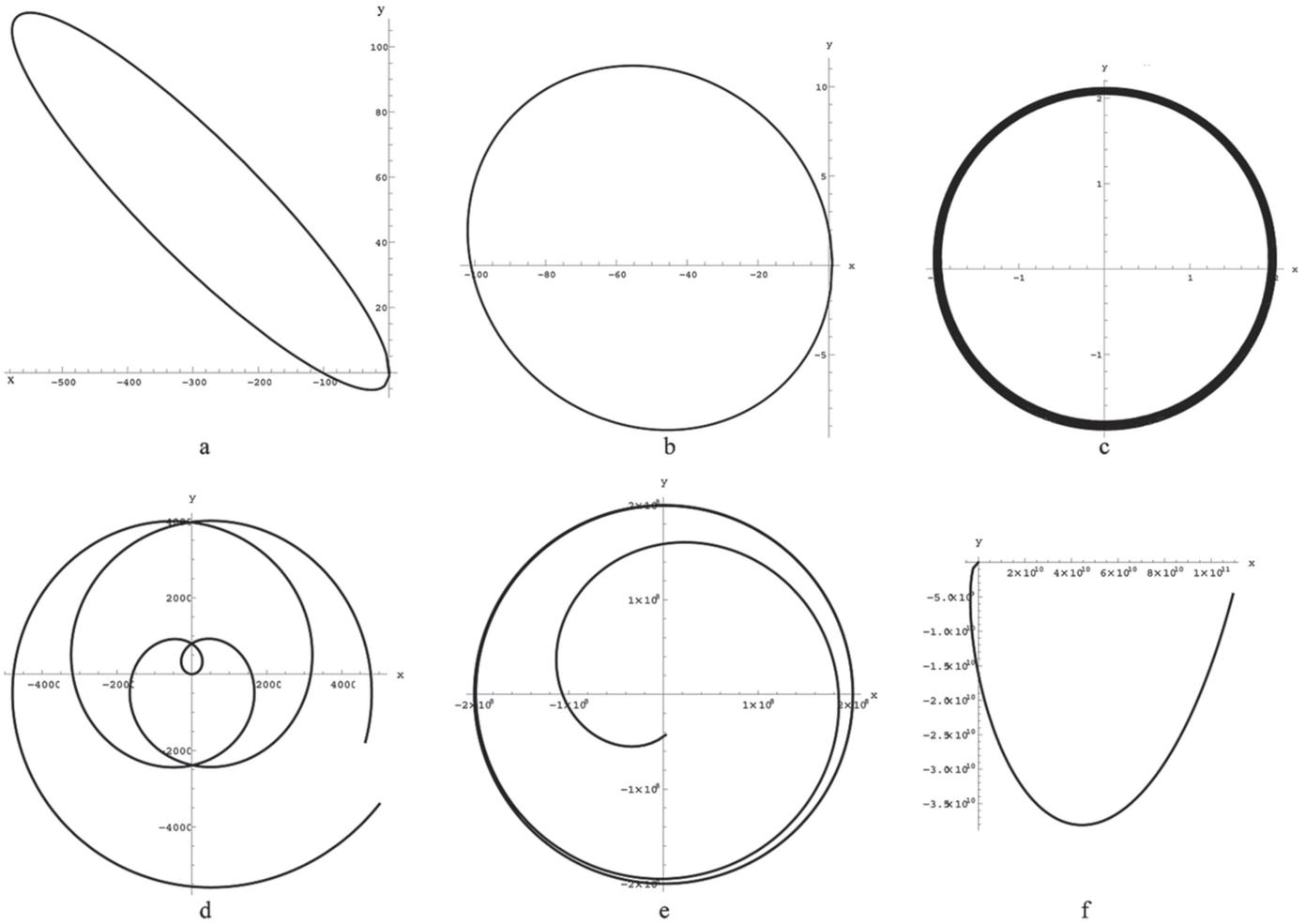

**Figure 11.** Examples of the trajectories of a test particle in the vicinity of the black hole from the point of view of the distant observer. The closed-space case, $\beta = 10^{16}$. (a) $L = 1$, $r(1.2) = 1.01$, $r'(1.2) = 0.09$, $t \in (1.2, 44750)$. (b) $L = 1$, $r(1.2) = 1.01$, $r'(1.2) = 0.01$, $t \in (1.2, 6087)$. (c) $L = 1.001$, $r(1.2) = 2.01$, $r'(1.2) = 0.0298$, $t \in (1.2, 7698)$. (d) $L = 2 \times 10^{-3}$, $r(1.01 \times 10^{16}) = 1.01$, $t \in (1.0099 \times 10^{16}, 1.01 \times 10^{16})$. (e) $L = 4.0 \times 10^{5}$, $r(10^{16}) = 1.42 \times 10^{7}$, $t \in (10^{16}, 2.5 \times 10^{16})$. (f) $L = 10^{4}$, $r(1.01 \times 10^{16}) = 1.02 \times 10^{7}$, $t \in (10^{16}, 1.01 \times 10^{16})$.

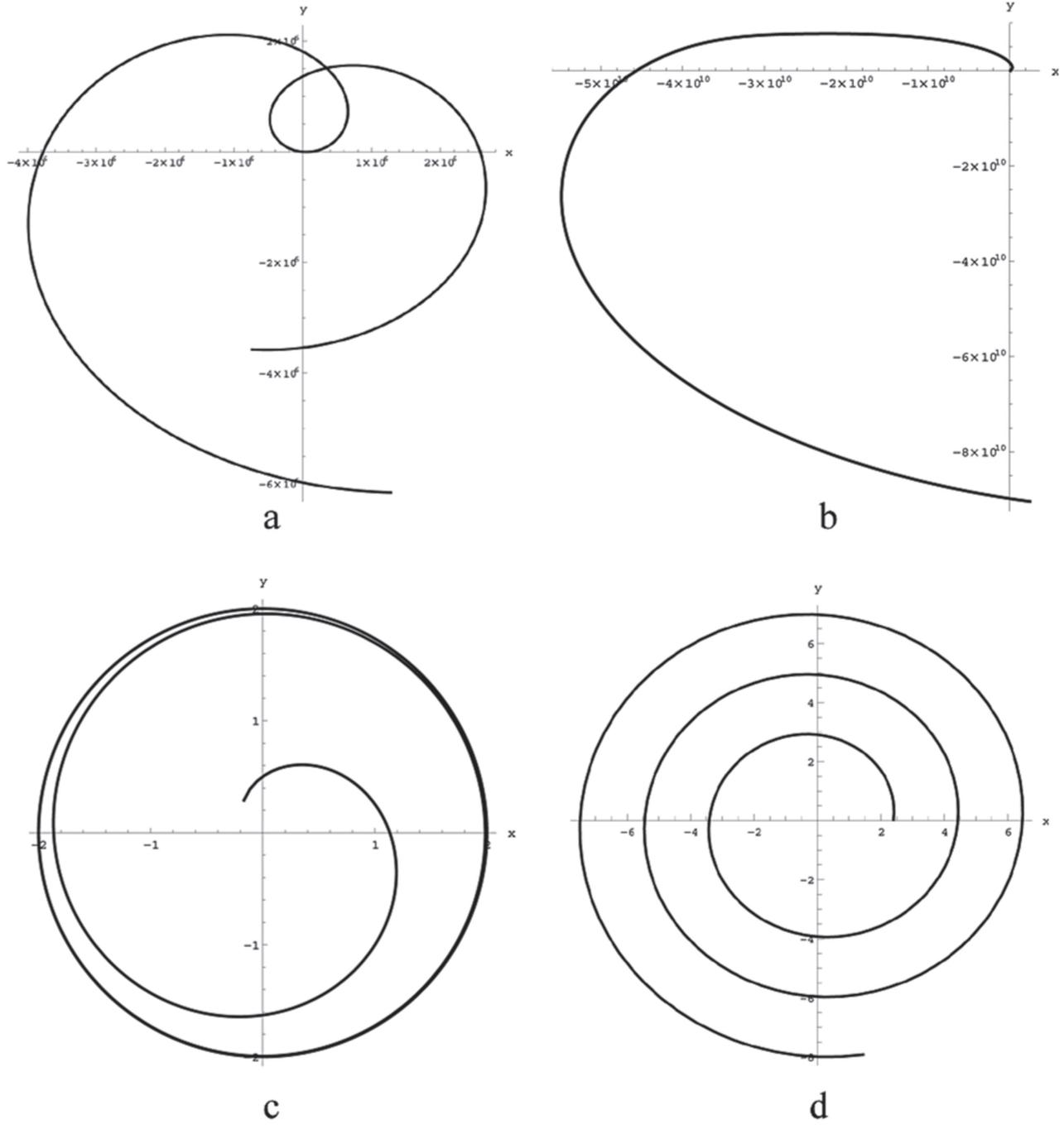

**Figure 12.** Examples of the trajectories of a test particle in the vicinity of the black hole from the point of view of the distant observer. The open-space case, $\beta = 10^{16}$. (a) $L = 5.015 \times 10^3$, $r(1.2033 \times 10^{16}) = 1.01 \times 10^5$, $r'(1.2033 \times 10^{16}) = 0.49653$, $t \in (1.20329 \times 10^{16}, 1.20330 \times 10^{16})$. (b) $L = 1.2 \times 10^5$, $r(1.201 \times 10^{16}) = 1.41 \times 10^8$, $t \in (1.200999 \times 10^{16}, 1.6 \times 10^{16})$. (c) $L = 1$, $r(1.01) = 2$, $t \in (1.01, 97.36563)$. (d) $L = 1$, $r(4 \times 10^{15}) = 2.4$, $t \in (4 \times 10^{15}, 4.0001 \times 10^{15})$.

From the figures, it can be seen that the behavior of the test particles near the black hole immersed in the cosmological medium in our model corresponds to what one may expect in the vicinity of the black hole. Besides, we have shown analytically that near the black hole every type of solution asymptotically tends to the Schwarzschild solution, while in the distant regions there is an expanding Friedmann-like background filled with homogeneous dust.

## 6. Conclusions

In this paper, we propose a way to construct the model of the cosmological black hole in the dust-filled universe on the basis of the exact solution to the Einstein equations of the Lemaître–Tolman–Bondi class. We have found such a solution as a particular case of the Tolman solution with the arbitrary functions chosen in a special way such that the solution includes both the Schwarzschild and Friedmann solutions as its natural limiting cases. We have analyzed the properties of the obtained solution for the three types of spatial curvature and built the R–T structure of the resulting spacetime, showing the horizons and singularities of the solution. We have demonstrated that in the center of symmetry of the obtained spacetime there is a region where the black hole is situated. The question of avoiding the shell-crossing in the model was solved. The trajectories of the test particles were built near the black hole for both comoving and distant observers. Although the metric for the distant observer was nonstatic, we have obtained the physically realistic picture of motion of the test particles around the black hole. From our analysis, it follows that within the second-order accuracy one will obtain the usual results, but cosmological corrections will appear in the next order. And so long as the obtained solution is exact the cosmological corrections to the Schwarzschild asymptotic may be found analytically up to any order of accuracy. All of the obtained results serve as convincing evidence in favor of the consistency of our exact solution for its future use in solving problems concerning the interplay between the cosmological expansion and local gravity, as well as other related problems.

The authors cordially thank Prof. Maria Korkina for useful discussions and Mgr. Andrej Babič for help in preparing the manuscript. This paper is supported by Grants of the Plenipotentiary Representative of the Czech Republic in JINR under Contract No. 192 from 05/04/2018 and the 3+3 Project No. 259 from 20/04/2017. Z.S. acknowledges the Albert Einstein Centre for Gravitation and Astrophysics supported by the Czech Science Foundation Grant No. 14-37086G. The authors acknowledge the Research Centre of Theoretical Physics and Astrophysics.